\documentclass{mn2e}
\usepackage{graphicx}

\begin{document}

\title{The asteroseismological potential of the pulsating DB white dwarf stars CBS\,114 and PG\,1456+103}
\author[Handler, Metcalfe \& Wood]
    {G. Handler,$^{1}$\thanks{E-mail: gerald@saao.ac.za} 
     T. S. Metcalfe,$^{2}$ M. A. Wood$^{3}$
	\and \\
$^{1}$ South African Astronomical Observatory, P.O. Box 9, Observatory 7935,
South Africa\\
$^{2}$ Theoretical Astrophysics Center, Institute of Physics and
Astronomy, Aarhus University, DK-8000 Aarhus C, Denmark\\
$^{3}$ Dept.~of Physics and Space Sciences \& SARA Observatory, Florida
Institute of Technology, Melbourne, FL 32901, USA\\
 }

\date{Accepted 2002 nnnn nn.
   Received 2002 nnnn nn;
   in original form 2002 nnnn nn}

\maketitle

\begin{abstract} 

We have acquired 65 h of single-site time-resolved CCD photometry of the
pulsating DB white dwarf star CBS\,114 and 62 h of two-site
high-speed CCD photometry of another DBV, PG\,1456+103. The pulsation
spectrum of PG\,1456+103 is complicated and variable on time scales of
about one week and could only partly be deciphered with our measurements.
The modes of CBS\,114 are more stable in time and we were able to arrive at
a frequency solution somewhat affected by aliasing, but still
satisfactory, involving seven independent modes and two combination
frequencies. These frequencies also explain the discovery data of the
star, taken 13 years earlier.

We find a mean period spacing of 37.1 $\pm$ 0.7\,s significant at the 98\%
level between the independent modes of CBS\,114 and argue they are due to
nonradial g-mode pulsations of spherical degree $\ell=1$. We performed a
global search for asteroseismological models of CBS\,114 using a genetic algorithm, 
and we examined the susceptibility of the results to the uncertainties of the
observational frequency determinations and mode identifications (we could
not provide $m$ values). The families of possible solutions are identified
correctly even without knowledge of $m$. Our optimal model suggests
$T_{\rm eff} = 21\,000$\,K and $M_* = 0.730~M_{\odot}$ as well as
$\log(M_{\rm He}/M_*) = -6.66, X_{\rm O} = 0.61$. This measurement of the
central oxygen mass fraction implies a rate for the $^{12}{\rm
C}(\alpha,\gamma)^{16}{\rm O}$ nuclear reaction near $S_{300}=180$\,keV\,b,
consistent with laboratory measurements.

\end{abstract}


\section{Introduction}

The helium-atmosphere DB white dwarf stars are very interesting from the
standpoint of stellar structure and evolution. The chemical evolution of
their atmospheres cannot be satisfactorily explained to date (see Shipman
1997 for a review). In particular, the presence of the so-called DB gap,
the absence of DB white dwarfs between temperatures of $\sim$30000 to 45000
K (Liebert 1986), is poorly understood. It is also not clear whether DBs
are mostly produced by single-star evolution or whether a significant
fraction originate from binary progenitors.

The examination of these problems can be aided by asteroseismology -- the
study of the interiors of pulsating stars via the analysis of their normal
mode spectra. Fortunately, a class of pulsating DB white dwarf stars
(hereinafter DBVs) exists, and their prototype, GD\,358, is one of the
classical examples for the successful application of asteroseismological
methods (Winget et al. 1994, Vuille et al. 2000).

Compared to other classes of pulsating star, the DBVs seem quite promising
candidates for asteroseismology: their mode spectra are shown or believed
to be rich, and pulsation theory for these stars is quite advanced and
well-tested. For instance, the central oxygen abundance of
the models has been shown to have a measurable effect on the pulsation
frequencies (Metcalfe, Winget \& Charbonneau 2001), and the possible
presence of a $^3$He/$^4$He transition zone due to chemical diffusion
has also been shown to produce a measurable effect (Montgomery, Metcalfe
\& Winget 2001), although recent observations have ruled out this
possibility (Wolff et al. 2002). In addition, the recently developed 
application of genetic-algorithm-based model-fitting to DBV
mode spectra (Metcalfe, Nather \& Winget 2000) allows an objective and
effective exploration of parameter space to find an optimal model for the
observations.

The application of these methods to DBVs other than GD\,358 is however
hampered by a lack of suitable observational data. Most of these stars are
rather faint (around 16$^{\rm th}$ magnitude), which requires the use of
2-m class telescopes to acquire suitable data with photomultiplier
detectors. However, time on these telescopes is generally oversubscribed and
long runs are difficult to obtain. With the advent of CCD
detectors and with the development of systems capable of acquiring
measurements with high time resolution, the DBVs are now in reach of smaller
telescopes. We have therefore started to acquire measurements aiming at
making as many DBVs as possible accessible to asteroseismology.

The pulsations of the DB white dwarf star CBS\,114 were discovered by
Winget \& Claver (1988, 1989), who reported multiperiodic oscillations
with peak-to-peak light variations up to 0.3 mag and with a time scale of
about 650\,s. PG\,1456+103 (hereinafter referred to as PG\,1456) was reported
as a DBV by Grauer et al. (1988). These authors reported multiperiodic
pulsations with dominant periodicities between 420 and 850 seconds and
maximum peak-to-peak amplitudes around 0.15 mag. As no further
observations of either of these stars have been published, we deemed it
worthwhile to obtain larger data sets to study the pulsations of CBS\,114
and PG\,1456 in more detail.

\section{Observations}

Most of our measurements were acquired as differential CCD photometry with
the 0.75-m telescope at the Sutherland station of the South African
Astronomical Observatory (SAAO), at which we had been allocated three
weeks of observing time.

We used the University of Cape Town CCD camera (O'Donoghue 1995). Its CCD
was operated in full-frame mode to acquire a maximum number of local comparison
stars. We used 2$\times$2 or 3$\times$3 prebinning depending on seeing,
which results in readout times of about 2--3 seconds. For the faint CBS
114 ($B\approx17.2$ as estimated from our CCD frames compared to PG
1456+103), we used integration times of 27 or 28 seconds to obtain one
frame per 30 seconds, and we used 20-second integrations for PG\,1456. No
filter was used to maximize the number of photons detected. On every clear
night, sky flatfields were taken during twilight.

In addition, we acquired measurements of PG\,1456 with the 0.9-m SARA
telescope at Kitt Peak National Observatory (Arizona), to reduce the
aliasing problem in the frequency analysis.   The CCD used is an Apogee
AP7p, with a back-illuminated SITe SIA-502AB 512$\times$512 detector.
The pixels are 24$\mu$m square, or 0.75 arcsec/pixel.  Read noise for
the camera is 12.2 electrons r.m.s., and the gain is 6.1 electrons per
ADU.  We used integration times of typically 30 seconds, and our
readout time was about 7 seconds.  Again, we took sky flats during each
twilight.
Our measurements are summarized in Table 1.

\begin{table}
\caption[]{Journal of the observations. The SARA measurements were 
obtained between JD 2451987--2451990; the remainder are SAAO data}
\begin{flushleft}
\begin{tabular}{cccccc}
\hline
\multicolumn{3}{c}{CBS\,114} & \multicolumn{3}{c}{PG\,1456+103}\\
 Run start & Length & N & Run start & Length & N \\
 JD-2450000 & h & & JD-2450000 & h &  \\
\hline
1940.421 & 3.32 & 403   & 1940.570 & 1.44 & 225\\
1941.400 & 3.74 & 450   & 1941.609 & 0.55 & 87\\
1942.394 & 4.00 & 446   & 1942.570 & 1.49 & 219\\
1943.446 & 1.30 & 143 \\
1944.427 & 3.40 & 404   & 1944.573 & 1.51 & 233\\
1945.397 & 4.25 & 484   & 1945.577 & 1.44 & 209\\
1946.386 & 4.17 & 435   & 1946.567 & 1.75 & 267\\
1954.379 & 0.68 & 80  \\
1955.359 & 4.20 & 382   & 1955.539 & 2.50 & 410\\
1956.357 & 4.30 & 513   & 1956.539 & 2.62 & 400\\
1957.353 & 4.25 & 495   & 1957.533 & 2.74 & 442\\
1959.346 & 4.38 & 464   & 1959.530 & 2.93 & 449\\
1960.346 & 4.30 & 511   & 1960.530 & 0.55 & 81\\
 & & & 1987.929 & 2.71 & 251\\
 & & & 1988.898 & 3.34 & 277\\
 & & & 1989.882 & 3.60 & 349\\
1996.249 & 4.17 & 497   & 1996.426 & 5.83 & 897\\
1997.239 & 4.32 & 511   & 1997.423 & 5.88 & 909\\
1998.235 & 4.36 & 520   & 1998.422 & 4.37 & 617\\
1999.355 & 1.37 & 156   & 1999.416 & 4.99 & 692\\
2001.319 & 1.38 & 107   & 2001.415 & 4.42 & 681\\
2002.252 & 2.77 & 324   & 2002.373 & 6.94 & 808\\
\hline
Total & 64.66 & 7325 & & 61.60 & 8503\\
\hline
\end{tabular}
\end{flushleft}
\end{table}

We started data reduction with the corrections for bias, dark counts (SARA
measurements only) and flatfield. Mean weekly flatfields were used for the
SAAO data sets and combined nightly flats for the SARA frames. Photometric
measurements on these reduced frames were made with the program MOMF
(Kjeldsen \& Frandsen 1992), which applies a combination of PSF and
aperture photometry relative to a user-specified ensemble of comparison
stars. No variability of any star other than the targets in the different
CCD fields was found, and the comparison star ensemble resulting in the
lowest scatter in the target star light curves was chosen.

The resulting differential light curves were corrected for differential
colour extinction and for correlations with seeing. For CBS\,114, it was
sometimes also found necessary to de-trend the data with (x, y) position
of the star on the chip because of flatfield errors. Residual
low-frequency trends in the data (not found to be coherent over the time
span of the observations, hence judged not to be intrinsic to the stars)
were removed by means of low-order polynomials.

Finally, the times of measurement were transformed to a homogeneous time
base. We chose Terrestrial Time (TT) as our reference for measurements on
the Earth's surface and applied a correction to account for the Earth's
motion around the solar system's barycentre. As this barycentric
correction varied up to $\pm$1\,s throughout a run, we applied it point by
point. Our final time base therefore is Barycentric Julian Ephemeris Date
(BJED). The reduced time series were subjected to frequency analyses; we
show example light curves in Figs.\,1 and 2. The typical rms scatter per
single data point is around 40 -- 50 mmag for CBS\,114, and for PG\,1456
about 15 mmag (SAAO data) and 11 mmag (SARA data).

\begin{figure}
\includegraphics[width=99mm,viewport=10 00 315 305]{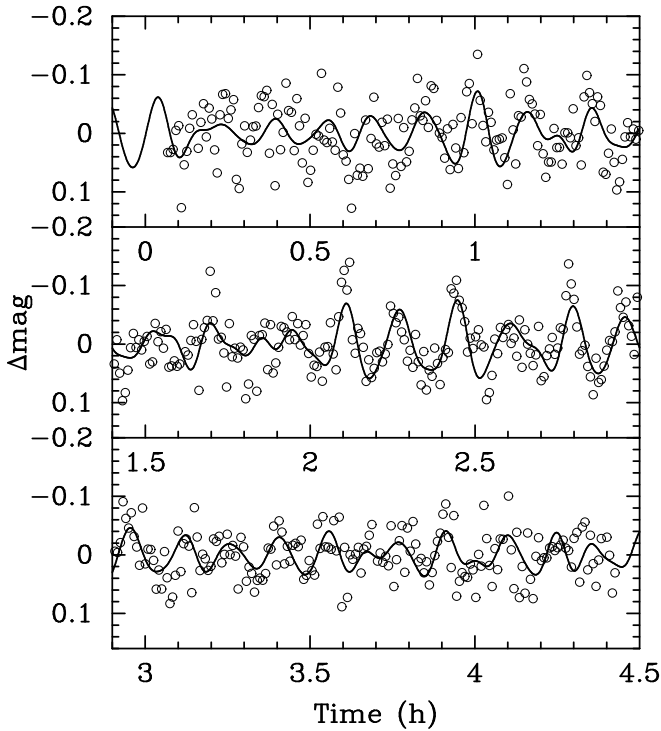}
\caption[]{An example light curve of CBS\,114; a multifrequency fit is
included as well. Although the data are not of impressive quality due to
the faintness of the star, the multiperiodic pulsations are clearly
visible.}
\end{figure}

\begin{figure}
\includegraphics[width=99mm,viewport=00 00 315 400]{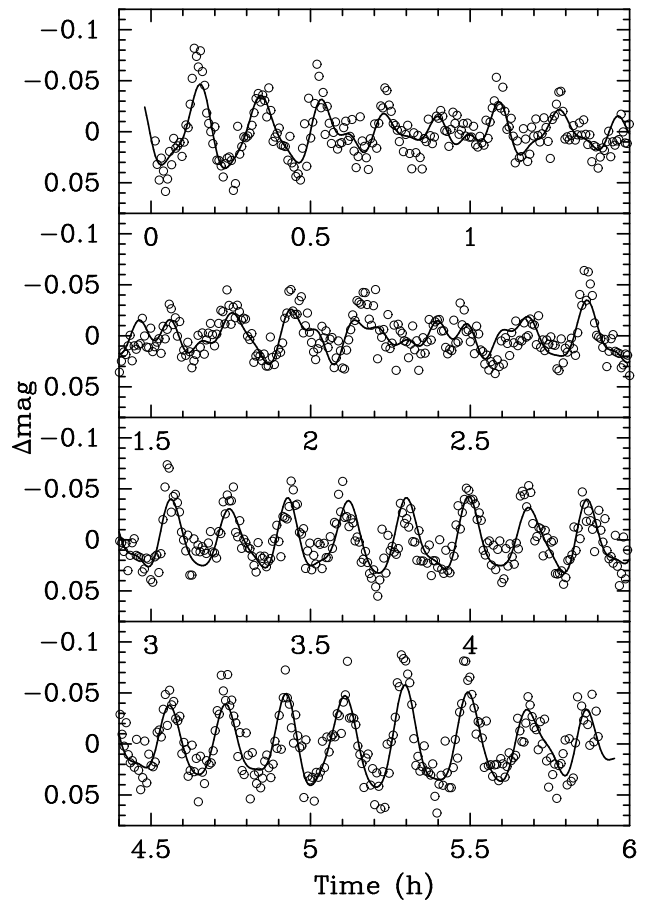}
\caption[]{An example light curve of PG\,1456+103 with a multifrequency
fit. As the star is 1.3 mag brighter than CBS\,114, the data are of much
better quality; the multiperiodic pulsations are readily visible as well.}
\end{figure}

\section{Frequency analysis}

Our frequency analysis was performed with the programme {\tt PERIOD 98}
(Sperl 1998). This package applies single-frequency power spectrum
analysis and simultaneous multi-frequency sine-wave fitting, but also has
some advanced options. In particular, it can be used to calculate optimal
solutions for multiperiodic signals including harmonic, combination, and
equally spaced frequencies, which are often found in the analysis of the
light curves of pulsating white dwarf stars.

\subsection{CBS\,114}

A frequency analysis of single-site data of a multiperiodic variable in
the presence of noise is difficult because of aliasing; extreme caution is
required. However, the structure and extent of our data set aids us in
this effort.

We started by calculating nightly amplitude spectra of our data. The
amplitude spectra of the individual nights were quite similar, with a
number of dominant peaks always occurring at similar frequencies. However,
some variations in the individual amplitudes were noted.

As the next step, we combined the three weekly data sets (which are very
similar in length and extent) and we computed their amplitude spectra as
well as that of all three weeks combined. They are shown in Fig. 3, which
shows six dominant structures between 1.4 and 2.6 mHz and some further
signals which could be combination frequencies.

\begin{figure}
\includegraphics[width=97mm,viewport=00 00 305 405]{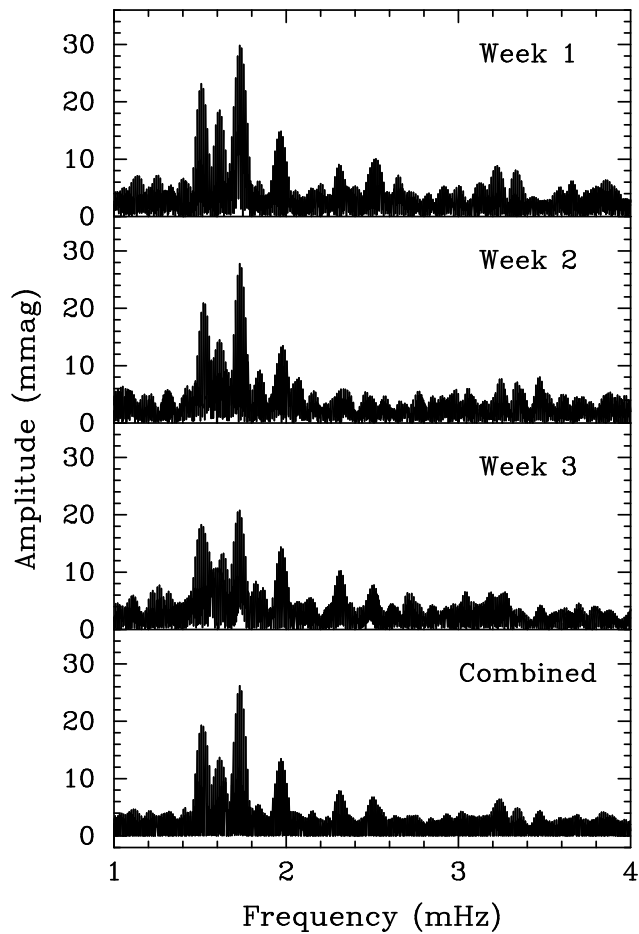}
\caption[]{Weekly and combined amplitude spectra of our CBS\,114 data. The
same peaks are always present, but with somewhat variable amplitudes.}
\end{figure}

Some trial prewhitening within the strongest features convinced us that
all of them are dominated by one signal each. Consequently, we determined
their frequencies by examining the combined, the weekly and the average of
the weekly amplitude spectra and by least-squares fitting to the light
curves. In this way, we could rule out a number of alias frequencies
because they gave unreasonable results if adopted. However, in some cases,
no unambiguous decision could be made.

Having determined the best set of frequencies, we refined their values by
calculating an optimal solution for the whole data set, prewhitened it
from the data, and calculated the amplitude spectrum of the residuals
(upper panel of Fig. 4). Because of the temporal changes in the weekly
amplitude spectra, it is not surprising that some residual mounds of
amplitude remained near the dominant frequencies.

\begin{figure}
\includegraphics[width=97mm,viewport=00 00 305 265]{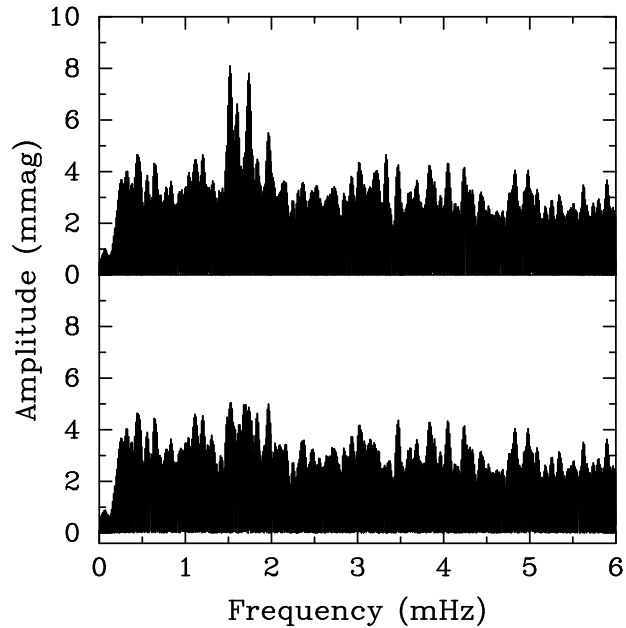}
\caption[]{Upper panel: residual amplitude spectrum after prewhitening the
best multifrequency solution for our measurements of CBS\,114, assuming
constant amplitudes and phases over the whole data set. Lower panel:
residual amplitude spectrum after prewhitening the same frequency
solution, but allowing for amplitudes and phases to vary over half-week
intervals.}
\end{figure}

We examined the frequencies of these features and their aliases and
performed trial prewhitening to check for multiplet structure possibly
present. No such evidence was found. We then fit our best frequencies
to the individual nights longer than 3\,h (assuring the dominant
structures are resolved) and determined their amplitudes, phases, and
corresponding error estimates (following Montgomery \& O'Donoghue 1999).
We found that significant amplitude and phase changes occurred over time
scales of 3 -- 4 days, and that either amplitude or phase variability
alone is not sufficient to explain the observations; both are present.

As we cannot determine the cause of these variations, we treated them
phenomenologically and calculated our final multifrequency solution with
constant frequencies as determined before, but with amplitudes and phases
variable over half-week periods. In this way, we can determine ranges of
the amplitudes of the dominant signals, and most of the residual mounds in
the amplitude spectrum are thus removed (lower panel of Fig.\,4). In
Table\,2, we list the best multifrequency solution determined with this
method.

\begin{table}
\caption[]{Multifrequency solution for CBS\,114. For frequencies labeled
with a minus sign, the negative daily alias ($f_i - 11.60\mu$Hz) may be
the correct frequency, and for frequencies labeled with a plus sign, the
positive daily alias ($f_i + 11.60\mu$Hz) cannot be ruled out as the
correct value. Amplitudes are listed in the ranges they assumed
during the observations.}
\begin{flushleft}
\begin{tabular}{ccccc}
\hline
ID & Freq. & Period & Amp. (2001) & Amp. (1988)\\
 & ($\mu$Hz) & (s) & (mmag) & (mmag) \\
\hline
$f_1$ & 1518.75$^-$ & 658.43 & 16--33 & 33 \\
$f_2$ & 1613.12 & 619.91 & 10--17 & 15 \\
$f_3$ & 1729.73 & 578.13 & 22--37 & 18 \\
$f_4$ & 1969.60 & 507.71 & 11--17 & $<5$\\
$f_5$ & 2306.38$^+$ & 433.58 & 4--12 & 13\\
$f_6$ & 2509.89$^-$ & 398.42 & 4--10 & 9\\
$f_1+f_3$ & 3248.48 & --- &  5--7 & $<7$\\
$f_2+f_3$ & 3342.85 & --- & 2--8 & $<6$\\
$f_7$ & 1835.64$^{+-}$ & 544.77 & $<4.2$ & 13 \\
\hline
\end{tabular}
\end{flushleft}
\end{table}

\subsubsection{Re-analysis of Winget \& Claver's data}

Although the measurements by Winget \& Claver (1988, 1989) only comprise
5.2\,h of data obtained during two nights, we can still use them for a
comparison with our results. The amplitude spectrum of these measurements
(upper panel of Fig.\,5) shows dominant features at frequencies agreeing
very well with the ones determined from our data. Consequently, we fitted
the first six frequencies from Table 2 to these measurements and noticed
that they were consistent with the older data. We therefore determined
corresponding frequencies, amplitudes and phases, subtracted this fit from
the data and calculated a residual amplitude spectrum, which is shown in
the lower panel of Fig.\,5, revealing the presence of yet another signal.

\begin{figure}
\includegraphics[width=97mm,viewport=00 00 305 265]{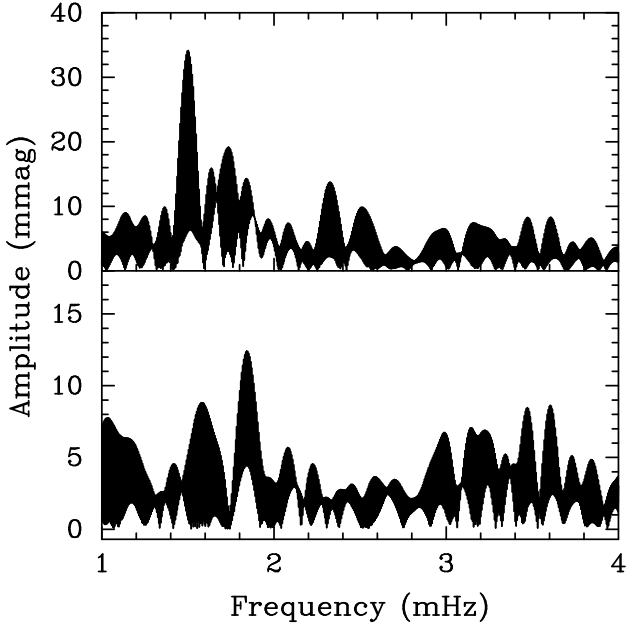}
\caption[]{Upper panel: amplitude spectrum of the measurements by  Winget
\& Claver (1988, 1989). Lower panel: residual amplitude spectrum after
prewhitening the first six frequencies determined from our observations.}
\end{figure}

A variation at this frequency is also present in our new observations,
although its amplitude is below our detection threshold. Consequently, 
we include it in our final frequency solution for the combined data sets
(Table 2). We note some more suspected signals in the lower panel of
Fig.\,5, but due to the small amount of data we do not push the analysis
further. We have therefore detected altogether seven independent pulsation
frequencies and two combination signals in the available light curves of
CBS\,114.

\subsection{PG\,1456}

Due to our two-site coverage of the light variations of PG\,1456, the
frequency analysis should be comparatively easy. However, initial trials
showed that this star's pulsational amplitudes and frequencies are also
somewhat variable, and as our light curves from the first two weeks are
rather short, they complicate the analysis if the total data set is used.
Consequently, we used the JD 2451987 - JD 2452002 two-site measurements as
our primary data set for the frequency analysis; the remaining data were
used for comparison purposes and consistency checks. Fig.\,6 shows some
prewhitening steps in these two main subsets of data.

\begin{figure}
\includegraphics[width=95mm,viewport=02 00 305 665]{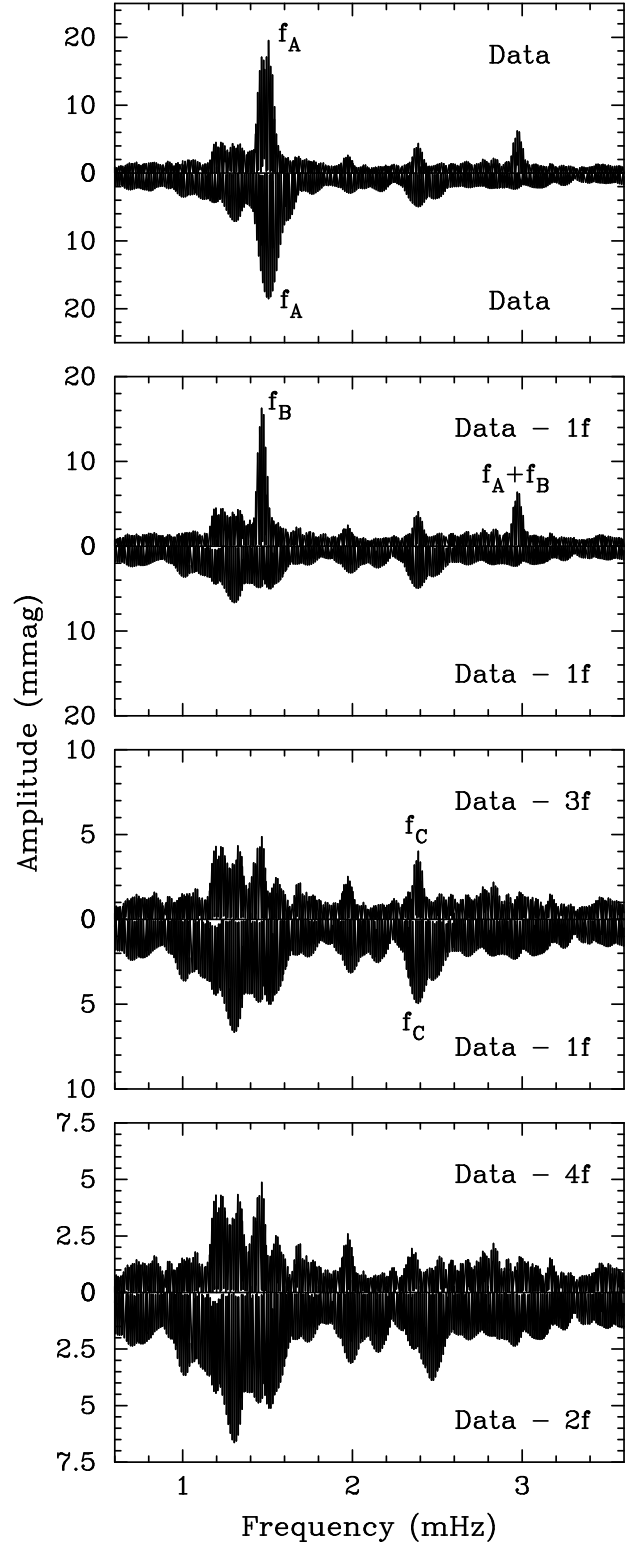}
\caption[]{Amplitude spectra of our measurements of PG\,1456. The upper
halves of the panels are for the two-site JD 2451987 - JD 2452002 data,
whereas the lower, mirrored, halves originate from the preceding
single-site observations. Uppermost panel: original data. Second panel:
data after prewhitening of the dominant frequency. Another frequency and a
combination are detected only in the two-site data. Third panel: another
mode is found in all the data. Lowest panel: the residual amplitude
spectra; more signals are clearly present.}
\end{figure}

A signal ($f_B$) not present in the early SAAO observations reached
considerable amplitude in the combined two-site data obtained about 7
weeks later; we note that this is not due to an aliasing problem. As a
matter of fact, the amplitude of this signal was still increasing during
the latter measurements. It had about 12 mmag in the SARA data, but 18
mmag in the last set of SAAO measurements, in which it also showed a
significant phase variation. A combination frequency ($f_A+f_B$) not
present initially also appeared in the two-site data. On the other hand,
the signal $f_C$ was found to be constant in amplitude and phase over the
whole data set, and $f_A$ was constant in amplitude but showed a slight
frequency shift of about 0.4 $\mu$Hz between our two subsets of data.

The residual amplitude spectra in the lowest panel of Fig.~6 still show a
number of interesting peaks, several of which may be real. To estimate the
richness of the frequency spectrum of PG\,1456, we performed further
prewhitening in the two-site data (still checking with the early SAAO
measurements), taking the temporal variations of $f_B$ into account. The
resulting amplitude spectra are shown in Fig.~7.

\begin{figure}
\includegraphics[width=99mm,viewport=02 00 305 350]{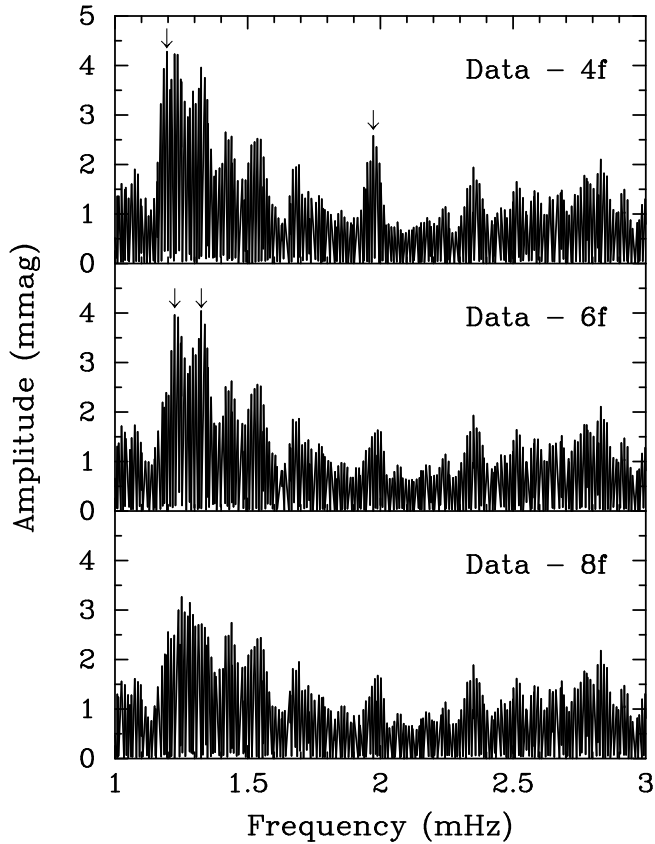}
\caption[]{Prewhitened amplitude spectra of our measurements of PG\,1456.
The arrows in the individual panels indicate further possible signals
which are prewhitened in the following panels.}
\end{figure}

It becomes clear that the light variations of PG\,1456 are quite
complicated and that our data set is not sufficient for finding all the
frequencies present. A preliminary multifrequency solution is given in
Table 3.

\begin{table}
\caption[]{Preliminary multifrequency solution for PG\,1456. Whereas we are
sure about the frequencies $f_A$ to $f_C$, the other values may be affected by
aliasing. Amplitudes are given for both the two-site measurements (left
column) and the preceding single-site SAAO data. Because of amplitude and
frequency variability, error estimates are only given for amplitudes of
apparently constant signals.}
\begin{flushleft}
\begin{tabular}{ccccc}
\hline
ID & Freq. & Period & Amp. & Amp. \\
 & ($\mu$Hz) & (s) & (mmag) & (mmag) \\
\hline
\multicolumn{5}{c}{\bf Certain detections}\\
$f_A$ & 1505.9 & 664.1 & 18.9 $\pm$ 0.3 & 18.5 $\pm$ 0.6\\
$f_B$ & 1465.4 & 682.4 & 11.7 -- 18.4 & $---$ \\
$f_A+f_B$ & 2971.3 & $---$ & 5.7 -- 6.8 & $---$ \\
$f_C$ & 2387.9 & 418.8 & 4.0 $\pm$ 0.3 & 5.0 $\pm$ 0.6\\
\hline
\multicolumn{5}{c}{\bf Possible further signals}\\
 & 1195.9 & 836.2 & 4.4 $\pm$ 0.3 & 4.4 $\pm$ 0.6 \\
 & 1973.4 & 506.7 & 2.7 $\pm$ 0.3 & $---$ \\
 & 1224.8 & 816.5 & 3.9 $\pm$ 0.3 & $---$ \\
 & 1324.9 & 754.8 & 3.9 $\pm$ 0.3 & $---$ \\
\hline
\end{tabular}
\end{flushleft}
\end{table}

A few comments are still necessary. Signals in the region around
1970\,$\mu$Hz are present in both data sets with similar amplitude
(e.g.,
the 1973.4\,$\mu$Hz signal in Table 3), but their frequencies are
different by 5\,$\mu$Hz, even taking aliasing into account. The dominant
frequency region in the residual single site-data is around 1304\,$\mu$Hz,
but neither this frequency nor its aliases are present in the two-site data. It seems
that only intensive multisite observations will make a good understanding
of the frequency spectrum of PG\,1456 possible.

The periods we determined are in very good agreement with those
detected by Grauer et al. (1988). Regrettably, a detailed comparison of
these periods is impossible as the discovery data are no longer available
(Grauer, private communication).

\section{Interpretation of the observational results}

\subsection{CBS\,114}

The independent pulsation periods of CBS\,114 as listed in Table 2 seem to
have similar spacing. We therefore searched for and determined their mean
spacing by calculating the Fourier power spectrum of the period values
with unit amplitude (Handler et al. 1997), which we show in Fig.~8. A mean
period spacing (significant at the 98\% level) is indicated, amounting to
37.1$\pm$0.7\,s. This result is corroborated by a Kolmogorov-Smirnov test
(not shown).

\begin{figure}
\includegraphics[width=97mm,viewport=02 05 305 180]{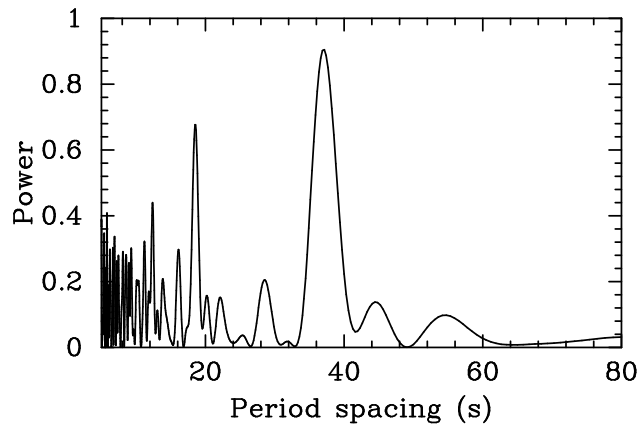}
\caption[]{Search for a preferred period separation within the independent
signals in the light curves of CBS\,114. A spacing of 37.1\,s and its
harmonics dominate this diagram.}
\end{figure}

We interpret this preferred period separation as a sign of the excitation
of a number of pulsational radial overtones with the same spherical degree
$\ell$. Some small deviations around its mean value (see Fig. 9 in
Sect.\,5) are an indication of mode trapping. The presence of a preferred
period separation is predicted by asymptotic theory (Tassoul 1980), and
was already observed within e.g. the $\ell=1, m=0$ modes of GD\,358 (Winget
et al. 1994, Vuille et al. 2000), for which a mean period spacing of
39.2\,s was found. In addition, Bradley \& Winget (1994) showed that the
systematic deviations from this mean spacing were due to mode trapping.

We cannot directly identify the $\ell, m$ values of the modes we observed
in CBS\,114 because of the lack of any rotational $m$-mode splitting in our
frequency spectra. However, it seems very likely that the independent
modes of CBS\,114 are all $\ell=1$: if we saw a mixture of $\ell$ values,
we would not expect to find such a significant mean period spacing.
Therefore the modes we observed must originate exclusively (or at least
predominantly) from the same $\ell$. They should be $\ell=1$ or 2 as
the effects of geometrical cancellation (Dziembowski 1977) are expected to
render modes of higher $\ell$ photometrically undetectable.

The mean period spacing between consecutive overtones of a pulsating white
dwarf star is a measure of its mass (Kawaler 1987). If the modes of CBS
114 were $\ell=1$, their mean period spacing would be consistent with the
star being a bit more massive than GD\,358. However, if the modes were all
$\ell=2$, CBS\,114 would need to have a mass below 0.3$M_{\sun}$ (see
Bradley, Winget \& Wood 1993). Such a low mass is inconsistent with the
spectroscopic gravity of the star (Beauchamp et al. 1999) and would be
quite unusual considering the mass distribution of the DB white dwarf
stars (Beauchamp et al. 1996).

Based on these arguments, we suggest that CBS\,114 is a DBV star pulsating
predominantly in nonradial g-modes of spherical degree $\ell=1$. We note
that its individual pulsation periods, especially the longer period
modes, are very similar to those of GD\,358.

\subsection{PG\,1456}

The interpretation of the frequency spectrum of PG\,1456 is more difficult.
A search for a mean period spacing such as in the previous section did not
give a significant result. This suggests that the modes we detected are
not of the same $\ell$ or that the star is a fast rotator. The modes of
$m \neq 0$ would then mask the possible patterns of equally spaced
periods.

About half of the indpendent periods we detected or suspect are similar to
periods of GD\,358, but the others are not. We are therefore unable to
interpret the mode spectrum of PG\,1456 at this stage. Extensive multisite
observations, e.g. with the Whole Earth Telescope (Nather et al. 1990),
are required to understand this star's pulsations.

\section{Asteroseismology of CBS\,114}

Using the optimization method developed by Metcalfe, Nather, \& Winget
(2000) and Metcalfe, Winget, \& Charbonneau (2001), we performed a global
search for the optimal model parameters to fit the 7 independent pulsation
periods of CBS\,114 listed in Table 2. The method uses a parallel genetic
algorithm to minimize the root-mean-square (rms) differences between the
observed and calculated periods for models with effective temperatures
($T_{\rm eff}$) between 20,000 and 30,000 K, total stellar masses ($M_*$)  
between 0.45 and 0.95 $M_{\odot}$, and helium layer masses with
$-\log(M_{\rm He}/M_*)$ between 2.0 and $\sim$7.0. This technique has been
shown to find the globally optimal set of parameters consistently among
the many possible combinations in the search space, but it requires between
$\sim$10 and 4000 times fewer model evaluations than an exhaustive search of
parameter-space to accomplish this (depending on the number of free 
parameters), with a failure rate $<10^{-5}$.

We assumed that all of the observed modes had spherical degree $\ell=1$
(as suggested in Sect. 4.1) and azimuthal order $m=0$. The latter
assumption can bias our determination of the {\it particular} set of model
parameters that produces the optimal fit to the data, but if the rotation
period is $\sim$1 day, any set of periods drawn from $m=(-1,0,+1)$ will
produce essentially the same overall picture of the parameter-space. We
demonstrated this general behavior by generating C-core fits to 100 data
sets for GD~358 that used randomly-selected $m$-components from those
identified in Winget et al.~(1994), and the $m=0$ values for $k=12,18$
(for which no triplet structure was found). In every case, the optimal set
of model parameters fell within the same families of good solutions that
were identified using the $m=0$ modes. Thus, when the spherical degree of
the modes is known, the model-fitting procedure correctly identifies the
{\it families} of possible solutions even when the values of $m$ are
unknown. Furthermore, we found that the optimal solutions fell into the
various families within parameter-space in proportion to the relative
fitness of that family when only $m=0$ modes were used for the fit. For
example, if a family has a peak fitness in the $m=0$ case that is
3$\sigma$ better than any other family, then the use of $m\neq 0$ modes
will identify an optimal solution in this same family with $\sim$50\%
probability. This means that if one of the families produces much better
solutions than the others, we can be reasonably confident that we have at
least identified the correct family of solutions.

\subsection{Carbon-core fit}

Assuming a pure C core extending to a fractional mass of $0.95~m/M_*$, the 
optimal set of model parameters found by the genetic algorithm for CBS\,114
were
$$
T_{\rm eff}=24,600~{\rm K},\ 
M_* = 0.655~M_{\odot},\ 
\log(M_{\rm He}/M_*) = -3.96,
$$
with rms period residuals $\sigma_P = 1.02~s$. The observed and calculated
periods are shown in the top panel of Fig. 9 plotted against the
deviations from the mean period spacing (dP), which were calculated using
the same set of periods in both cases. Each point in this representation
of the data is independent of the others, unlike a period spacing diagram
using $\Delta P$ ($\equiv P_{k+1} - P_k$; cf. Bradley \& Winget 1994).
Note that the genetic algorithm only fits the periods of the pulsation
modes, and the agreement between the deviations from the mean period
spacing is simply a reflection of the overall quality of the match.

\begin{figure}
\includegraphics[width=80mm]{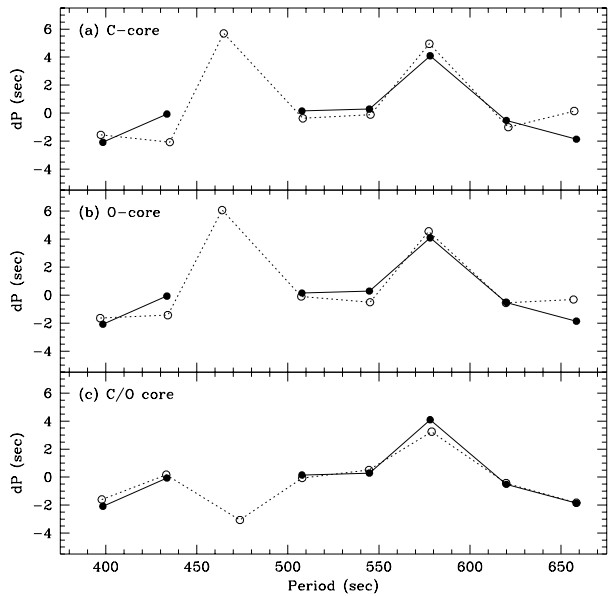}
\caption[]{The observed periods of CBS\,114 (solid) and the optimal model
periods found by the genetic algorithm (open) plotted against the
deviations from the mean period spacing, assuming a core of (a) pure C (b)
pure O, and (c) mixed C/O (see text for complete details).}
\end{figure}

As noted in Table 2, the daily aliases of several of the identified modes
in CBS\,114 cannot be ruled out entirely. We can assess the impact of this
uncertainty on the final set of optimal model parameters by repeating the
fitting procedure using one or more of the alias periods. However, since
the genetic algorithm fitting method with 3 free parameters is only about
ten times more efficient than calculating the entire grid of models, it is
better to calculate the pulsation periods of all $10^6$ models if we
intend to repeat the procedure more than a few times. As there are four
periods with a total of five possible aliases, we chose to calculate the
entire grid of models. This allows us to check the answer that resulted
from the genetic algorithm fit, and will enable us to generate C-core fits
very quickly in the future for any additional data sets on DBV white dwarf
stars.

All C-core models with rms period residuals smaller than 2.22 seconds are
shown in the top half of Fig.10, which includes front and side views of
the entire search space. Each point in the left panel corresponds to a
point in the right panel, and the darkness of a point indicates the
relative quality of the match with the observations. Black points are
within $\Delta\sigma_P=0.03~s$ of the optimal model, and the four
progressively lighter shades of grey correspond to models within 3, 10,
25, and 40 times this difference. The parameter-correlations explained by
Metcalfe, Nather, \& Winget (2000) are clearly visible, causing the good
models to fall along lines in the plot rather than on a single point.

\begin{figure}
\includegraphics[width=80mm]{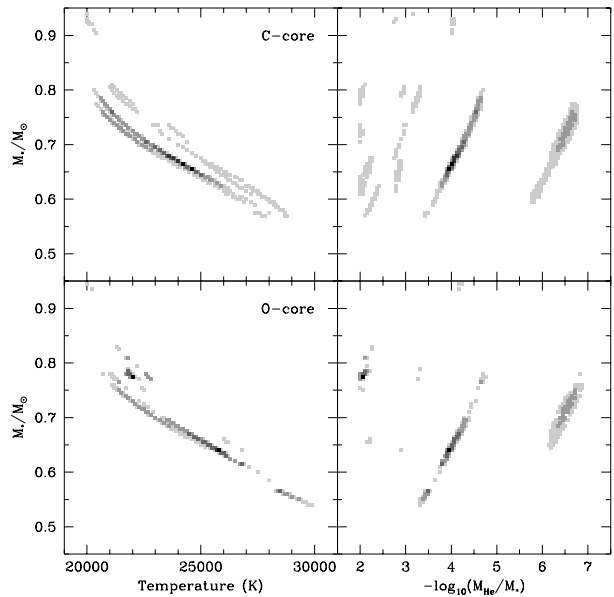}
\caption[]{Front and side views of the search space assuming an C-core
(top panels) and an O-core (bottom panels). Square points mark the
location of models that yield a reasonable match to the periods observed
for CBS\,114. The darkness of a point indicates the relative quality of
the match (see text for complete details).}
\end{figure}

As expected, the optimal parameters from comparison of the observations to
the complete grid of models were identical to those found using the
genetic algorithm method. In fact, the identified periods led to lower rms
residuals than any set of periods that included one of the possible
aliases. Of the combinations with two periods replaced by their aliases,
fits with lower residuals were achieved in three cases: (1)
$f_1\rightarrow f_1^-$ and $f_7\rightarrow f_7^+$, (2) $f_1\rightarrow
f_1^-$ and $f_7\rightarrow f_7^-$, and (3) $f_6\rightarrow f_6^-$ and
$f_7\rightarrow f_7^-$. When either three or four periods were replaced
with aliases, only one combination led to lower residuals: $f_1\rightarrow
f_1^-$, $f_5\rightarrow f_5^+$, and $f_7\rightarrow f_7^+$. Note that
aliases of $f_1$ and $f_7$ appear in most of these alternative period
lists, implying that they are relatively important to the outcome of the
fit. Although lower residuals were possible using these various
combinations of two or more alias periods, the optimal models in every
such case are too massive and too cool to be reconciled with the spectral
line fits for CBS\,114 by Beauchamp et al.~(1999).

\subsection{Oxygen-core fit}

The optimal set of model parameters found by the genetic algorithm for a
pure O core extending to $0.95~m/M_*$ had a mass that was inconsistent
with the measurements of Beauchamp et al.~(1999) and a helium layer mass
at the edge of our search range, near the theoretical limit where nuclear
burning will occur at the base of the envelope. The second-best model had
residuals only $\sim$0.03 seconds higher and was similar to the optimal
C-core model.
The parameters of this model were
$$
T_{\rm eff}=25,800~{\rm K},\
M_* = 0.640~M_{\odot},\
\log(M_{\rm He}/M_*) = -3.96,
$$
with rms period residuals $\sigma_P = 0.91~s$. The middle panel of Fig. 9 
shows the calculated periods of this model along with the observations, 
plotted against the deviations from the mean period spacing. The fit is 
only slightly better than the C-core model, and is qualitatively similar.

The O-core models that were calculated by the genetic algorithm during the
optimization process with rms period residuals smaller than 2.08 seconds
are shown in the bottom half of Fig.\,10. Although this is not a complete
sampling of the parameter-space, it is heavily sampled in regions where
models produce better than average residuals. Again, the shade of each
point indicates the relative quality of the match---black points are within
$\Delta\sigma_P=0.03~s$ of the optimal model, and the shades of grey are at 3,
10, 25, and 40 times this difference. Note that because the optimal O-core
model has residuals $\sim$0.1 seconds lower than the optimal C-core model,
the darkest grey points in the bottom half of Fig. 10 are actually better
than the the black points in the top half.

\subsection{C/O-core fit}

Since it may seem dubious to fit the 7 observed periods of CBS\,114 using a
model with five free parameters, we proceed cautiously. To allow a
systematic exploration of models with various internal C/O profiles,
Metcalfe, Winget, \& Charbonneau (2001) used a simple parameterization that
explored a general class of profiles similar to those used by Bradley,
Winget, \& Wood (1993). The parameterization fixes the oxygen mass fraction
to its central value ($X_{\rm O}$) out to some fractional mass ($q$) where
it then decreases linearly in mass to zero oxygen at $0.95~m/M_*$.

Using this method, the optimal model parameters for the observed pulsation 
periods of CBS\,114 were
$$
T_{\rm eff}=21,000~{\rm K},\
M_* = 0.730~M_{\odot},
$$ 
$$
\log(M_{\rm He}/M_*) = -6.66,\ 
X_{\rm O} = 0.61,\ 
q = 0.51,
$$
with rms period residuals $\sigma_P = 0.43~s$. All models within Delta
sigma = 0.03 s of this fit had the same mass, temperature, and helium
layer mass at our sampling resolution of 0.005 $M_{\sun}$, 100 K, and 0.05 
dex respectively. Only models with a central oxygen mass fraction within 
Delta $X_O = \pm 0.01$ and the optimal value of q had rms residuals within 
this range.

The calculated periods of this model are shown in the bottom panel of
Fig.\,9 along with the observed periods, plotted against the deviations
from the mean period spacing. To evaluate whether or not this fit is
better by an amount that justifies the addition of two free parameters, 
we
can use the Bayes Information Criterion (BIC, following Koen \& Laney
2000):
$$
{\rm BIC} = N_p \left( \log N_{\rm obs} \over N_{\rm obs} \right) 
+ \log\sigma^2
$$
where $N_p$ is the number of free parameters, $N_{\rm obs}$ is the number
of observed periods, and $\sigma$ is the rms period residual of the
optimal fit. The value of BIC must be lower for a decrease in $\sigma$ to
be considered significant. Our best 3-parameter fit was the O-core model,
which leads to a value of ${\rm BIC}=0.28$. This leads us to expect the
residuals of a 5-parameter fit to decrease to 0.69 seconds without being
considered significant. The rms residuals of our 5-parameter fit are
substantially lower than 0.69 seconds, so the addition of the extra
parameters seems to be justified, and the C/O fit is significantly better
than the O-core model.

Even so, we might worry that our optimal model parameters may be less
accurate than the values obtained for GD~358, due to the 
smaller number of
observed pulsation periods. To determine the magnitude of any systematic
uncertainties due to the smaller number of observed modes, we performed a
new fit to the pulsation periods of GD~358, but using only
the 7 modes corresponding to those observed in CBS\,114: 
$k=8,9$ and 11-15.  
When we compared the result of this fit with the one using all 11 modes
observed in GD~358, we found only small shifts to the optimal model
parameters:
$$
\Delta T_{\rm eff}=-500~{\rm K},\
\Delta M_* = +0.015~M_{\odot},
$$$$
\Delta \log(M_{\rm He}/M_*) = -0.05,\
\Delta X_{\rm O} = -0.02,\
\Delta q = -0.01.
$$
By analogy, we might expect our 5-parameter fit to CBS\,114 to represent a
slight underestimate of the temperature and overestimate of the mass. Both
potential biases help to explain part of the discrepancy between the mass
and temperature of our optimal model for CBS\,114 and the values inferred from the
spectroscopic analysis of Beauchamp et al.~(1999) [$T_{\rm eff}=23,300$;
$\log\ g=7.98$]. Additional differences are expected since we have used
different mixing-length parameters (ML3) than Beauchamp et al.  
(ML2/$\alpha=1.25$). Metcalfe, Salaris, \& Winget (2002) quantified the
offsets between fits to GD~358 using ML3 and ML2, and found them to be
approximately the same size as the shifts due to the smaller number of
data points. Unfortunately, CBS\,114 does not have a published parallax, so
an independent constraint on the mass and temperature from the luminosity
is not presently available.

\section{Conclusions}

We have presented time-resolved CCD photometry of the pulsating DB white
dwarf stars CBS\,114 and PG\,1456+103. Our data, obtained with
telescopes
of only 0.75\,m and 0.9\,m apertures, are sufficient for the extraction of useful
asteroseismological information even for a star as faint as CBS\,114 ($B
\approx 17.2$). We have also shown that it is possible to understand the
mode spectra of multiperiodic pulsating white dwarf stars from single-site
observations in suitable cases (see Handler 2001 for another example).
In other cases, like PG\,1456, this is not possible, and worldwide
observing campaigns, e.g. with the Whole Earth Telescope, are necessary.

The frequency analysis of our measurements of CBS\,114 resulted in the
discovery of a mean period spacing of $37.1 \pm 0.7$\,s in the
independent modes of the star. Although no convincing evidence for
rotational splitting within the modes was detected, we argued that the
star pulsates in nonradial $\ell=1$ g-modes.

We used our data set to check the sensitivity of the genetic-algorithm-based
model-fitting method (Metcalfe et al. 2000) to
artifacts of data analysis, like aliasing problems and our lack of
assignments of the azimuthal order $m$ to the observed modes.
Encouragingly, our results are hardly affected by these
shortcomings.

Most significantly, the genetic algorithm found a C/O fit to the pulsation
periods of CBS\,114 that proved to be significantly better than the pure-core
fits, with an accuracy comparable to a similar fit to GD~358. The
optimal mass and central oxygen mass fraction from this fit can provide an
independent measurement of the astrophysically-important rate for the
$^{12}{\rm C}(\alpha,\gamma)^{16}{\rm O}$ reaction, as was recently done
for GD~358 (Metcalfe, Salaris, \& Winget 2002). In general, higher mass
white dwarfs are expected to have a lower central oxygen mass fraction,
because the 3$\alpha$ rate rises faster with increasing density than the
$^{12}{\rm C}(\alpha,\gamma)^{16}{\rm O}$ rate. Our optimal model for CBS
114 has a higher mass and a lower central oxygen mass fraction than the
optimal model for GD~358 (Metcalfe, Winget, \& Charbonneau 2001), so it is
consistent with this general trend. A model of the internal chemical
profile with the same mass as our fit to CBS\,114 requires a rate for the
$^{12}{\rm C}(\alpha,\gamma)^{16}{\rm O}$ reaction near $S_{300} = 180$
keV b to produce a central oxygen mass fraction of 0.61 (M. Salaris,
private communication). This value is close to the rate derived from
recent high-energy laboratory measurements ($S_{300} = 165 \pm 50$ keV b;
Kunz et al. 2002). By contrast, the rate derived from the optimal model of
GD~358 by Metcalfe, Salaris, \& Winget (2002) was significantly higher
($S_{300} = 370 \pm 40$ keV b). This suggests either that presently
unknown sources of
systematic uncertainty in our models must affect the analysis of GD\,358
and CBS\,114 in different ways, or that the two stars have different
evolutionary origins, or both. An asteroseismological determination of the
central oxygen mass fraction for additional DBV white dwarfs will help us
to decide which of these scenarios is most likely.

\section*{ACKNOWLEDGEMENTS}

We thank Don Winget and Chuck Claver for permission to use their data of
CBS\,114, Maurizio Salaris for providing a new internal chemical
profile to match our optimal model and Mike Montgomery for carefully
reading a draft version of this paper.

\end{document}